\documentclass{fizik}%
\usepackage{multicol,epsfig}

\vol{27}
\pyear{2003}
\received{\today}
\author[Aharony, Entin-Wohlman, Imry]{
\textbf{Amnon AHARONY$^a$, Ora ENTIN-WOHLMAN$^a$, and Yoseph IMRY$^b$ }\\
\textit{$^a$ School of Physics and Astronomy, Raymond and Beverly
Sackler
Faculty of Exact Sciences,}\\
\textit{Tel Aviv University, Tel Aviv 69978,
ISRAEL}\\
\textit{$^b$ Department of Condensed Matter Physics, The Weizmann
Institute of Science, Rehovot 76100, ISRAEL}\\
}

\title{Phase measurements in Aharonov-Bohm interferometers}

\begin{document}
\maketitle

\begin{abstract}
In this paper we address measurements of the resonant quantum
transmission amplitude $t_{QD}=-i|t_{QD}|e^{i\alpha_{QD}}$ through
a quantum dot (QD), as function of the plunger gate voltage $V$.
Mesoscopic solid state Aharonov-Bohm interferometers (ABI) have
been used to measure the ``intrinsic" phase, $\alpha_{QD}$, when
the QD is placed on one of the paths. In a ``closed"
interferometer, connected to two terminals, the electron current
is conserved, and Onsager's relations require that the conductance
${\cal G}$ through the ABI is an even function of the magnetic
flux $\Phi=\hbar c\phi/e$ threading the ABI ring. Therefore, if
one fits ${\cal G}$ to $A+B\cos(\phi+\beta)$ then $\beta$ only
``jumps" between 0 and $\pi$, with no relation to $\alpha_{QD}$.
Additional terminals open the ABI, break the Onsager relations and
yield a non-trivial variation of $\beta$ with $V$. After reviewing
these topics, we use theoretical models to derive three results on
this problem: (i) For the one-dimensional leads, the relation
$|t_{QD}|^2 \propto \sin^2(\alpha_{QD})$ allows a direct
measurement of $\alpha_{QD}$. (ii) In many cases, the measured
${\cal G}$ in the closed ABI can be used to extract {\it both}
$|t_{QD}|$ and $\alpha_{QD}$. (iii) For open ABI's, $\beta$
depends on the details of the opening.  We present quantitative
criteria (which can be tested experimentally) for $\beta$ to be
equal to the desired $\alpha_{QD}$: the ``lossy" channels near the
QD should have both a small transmission and a small reflection.
 \keywords{interference in
nanostructures, Aharonov-Bohm interferometer, quantum dots,
resonant transmission.}
\end{abstract}

\section{Introduction and Review of Experiments}

Recent advances in the fabrication of nanometer scale electronic
devices raised much interest in the quantum mechanics of quantum
dots (QDs), which represent artificial atoms with experimentally
controllable properties \cite{review,book}. A flexible method to
construct mesoscopic QDs is based on the two dimensional electron
gas (2DEG), which exists in the planar interface between an
insulator and a semiconductor, with a metallic layer under the
insulator. Metallic electrodes, which are placed above the
semiconducting layer, create potentials on the 2DEG which restrict
the electrons to move only in parts of the plane \cite{kastner}.
The simplest QD geometry consists of a small bounded region, which
can bind electrons. This QD is connected via two one-dimensional
(1D) `metallic' leads to electron reservoirs. The coupling of each
lead to the QD is controlled by a potential barrier. The potential
on the QD itself, called the `plunger gate voltage', $V$,
determines the attraction of electrons to the QD, and thus also
the energies of electronic bound states on the QD. The simplest
experiments then measure the conductance ${\cal G}$ through the
QD, as function of $V$. The measured ${\cal G}$ shows peaks
whenever the Fermi energy $\epsilon_F$ of the electrons crosses a
bound state on the QD. Quantum mechanically, we should think of an
electronic wave, $e^{ikx}$, hitting the QD from the left. One then
ends up with a reflected wave, $r_{QD}e^{-ikx}$ and a transmitted
wave, $t_{QD}e^{ikx}$. The quantum information on the resonant
tunneling through the QD is contained in the {\it complex}
transmission amplitude, $t_{QD}=-i\sqrt{{\cal
T}_{QD}}e^{i\alpha_{QD}}$. It is thus of great interest to measure
both the magnitude $T_{QD}$ and the phase $\alpha_{QD}$, and study
their dependence on $V$.

\begin{figure}[htb]
\begin{center}
 \includegraphics[scale=0.75]{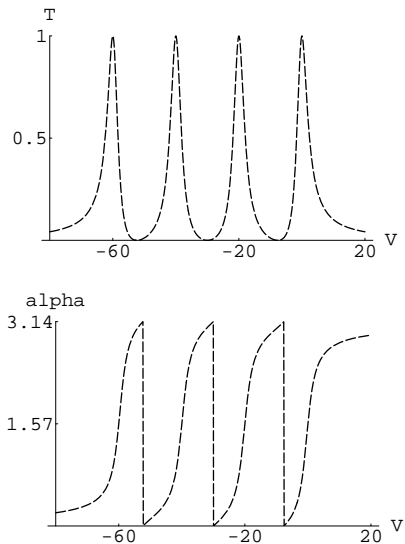}
\end{center}
\caption{Theoretical transmission ${\cal T}_{QD}$ and ``intrinsic"
phase $\alpha_{QD}$ for $N=4$ states on the QD, with ``gap"
$U=20J$, versus the gate voltage $V$ (in units of $J$). See Sec.
\ref{sec2} for details.} \label{ist1}
\end{figure}

Theoretically, the phase $\alpha_{QD}$ is particularly
interesting, given its relation to the additional electron
occupation in the system via the Friedel sum rule
 \cite{friedel,langreth}. This phase is also predicted to exhibit
interesting behavior e.g. near a Kondo-like resonance
\cite{hewson}. For a simple model of non-interacting electrons
with several equidistant bound state energies, theory yields the
magnitude and the phase as shown in Fig. \ref{ist1} (see below for
details): ${\cal T}_{QD}$ exhibits resonances at the bound state
energies, while $\alpha_{QD}$ exhibits an interesting variation
between 0 and $\pi$, growing gradually through each resonance, and
dropping sharply between consecutive resonances (here and in all
following graphs, we set $\alpha$ and $\beta$ at zero far below
the resonances). The resonant dependence of ${\cal T}_{QD}$ on $V$
has been confirmed by many experiments \cite{review,book}, which
measure the conductance and take advantage of the Landauer formula
\cite{landauer}, ${\cal G}=\frac{2e^2}{h}{\cal T}_{QD}$. However,
the experimental measurement of $\alpha_{QD}$ has only become
accessible since 1995 \cite{yacoby,schuster}, using the
Aharonov-Bohm (AB) interferometer \cite{azbel}. As explained
below, many experiments measure a phase (which we call $\beta$)
which 'oscillates' between 0 and $\pi$. However, the relation of
these measured values to the desired $\alpha_{QD}$ is not trivial.
This relation is one of the main topics of this review.

The simplest method to measure the phase of a wave is based on the
{\it two-slit interferometer} \cite{feynman}. In this geometry, a
coherent electron beam is split between two paths, going through
two slits, and one measures the distribution of electrons absorbed
on a screen behind the two slits. Assuming that each electron goes
through one of the slits only {\it once}, without any reflection
from the slits or from the screen, and assuming complete
coherence, the distribution of electrons on the screen is given by
${\cal T}=|t|^2$, where $t=t_1+t_2$ is the sum of the (complex)
amplitudes of the waves which went via the two slits.

In the two-slit AB interferometer, one adds a magnetic flux $\Phi$
in the area surrounded by the two electronic paths. Such a flux
creates a non-zero electromagnetic vector potential, ${\bf A}$,
even where the flux vanishes. With an appropriate choice of gauge,
the kinetic energy of the electron becomes $({\bf p}+e {\bf
A}/c)^2/(2m)$, where ${\bf p}$ is the electron momentum. As a
result, the wave function of the free electron which moves from
${\bf r}_1$ to ${\bf r}_2$ obtains an additional phase
$\phi_{12}=(e/\hbar c)\int_{{\bf r}_1}^{{\bf r}_2}{\bf A}({\bf
r})\cdot d{\bf r}$, where the integration is along the path of the
electron.
 Aharonov and Bohm \cite{AB} used this fact to predict that such a flux
 between the two paths would add a difference $\phi=e\Phi/\hbar c$
between the phases of the wave functions in the two branches of
the ring, yielding
\begin{equation}
t=t_1 e^{i\phi}+t_2. \label{2slit}
\end{equation}
(Gauge invariance allows one to attach the AB phase $\phi$ to
either branch). Writing $t_i=|t_i|e^{i \alpha_i}$, one thus has
\begin{equation}
{\cal T}=A+B\cos(\phi+\alpha), \label{T2slit}
\end{equation}
where $\alpha=\alpha_1-\alpha_2$. Assuming that one of the phases
can be varied experimentally (e.g. by placing a QD on one path and
changing its plunger gate voltage $V$), this `2-slit formula' can
then be used to deduce the dependence of the phase $\alpha$ on
external parameters (e.g. $V$).

Experiments using two-slit geometries for electron interference,
using electron microscopes, which confirmed the AB effect, have
been described in detail by Tonomura \cite{tonomura}. In the
present paper we concentrate on experiments which use {\it
mesoscopic devices}. A coherent flow of electrons requires that
the mean free path $L_\varphi$, over which scattering destroys the
electron's phase, should be larger that the sample size. This can
be achieved by going to low temperatures and by using small
samples. The first confirmation of the AB effect in a mesoscopic
system was done by Webb {\it et al.} \cite{webb}. They used a
small metal ring, which was connected (at two opposite points) to
electron reservoirs through two leads. Indeed, the conductance of
the ring showed a periodic dependence on the magnetic flux inside
the ring, $\Phi$, with a leading Fourier component at the period
$e/\hbar c$, as expected. However, this experiment did not allow a
variation of the relative phase $\alpha$, nor a detailed test of
the two-slit formula (\ref{T2slit}); specifically, the Fourier
analysis contained also higher harmonics.

\begin{figure}[htb]
\begin{center}
 \includegraphics[scale=0.75]{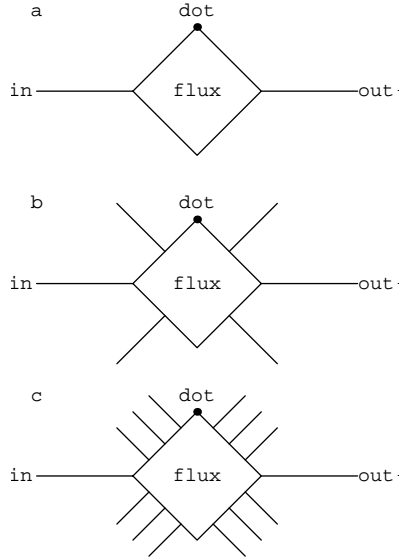}
\end{center}
\caption{Model for the AB interferometer: (a) Closed two-terminal
case, (b) Schematic picture of the six-terminal open
interferometer, (c) Model for the open interferometer.}
\label{ist2}
\end{figure}

The first attempt to vary the phase of the wave on one of the
paths was done by Yacoby {\it et al.} \cite{yacoby}. They used the
semiconducting QD system described above, in which the electrons
were also allowed to go via a `reference' path, parallel to the
path containing the QD (see Fig. \ref{ist2}a). Again, the measured
conductance was periodic in $\phi$, and the detailed dependence of
$\cal G$ on $\phi$ varied with the plunger gate voltage on the QD,
$V$. Far away from a resonance, this conductance could be fitted
to Eq. (\ref{T2slit}). However, closer to a resonance the data
seem to require more harmonics in $\phi$, e.g.
 of the form
\begin{equation}
{\cal T}=A+B\cos(\phi+\beta)+C\cos(2 \phi+\gamma)+\ldots,
\label{fit}
\end{equation}
with the conventions $B,~C>0$. Surprisingly, the fitted phase
$\beta$ did not vary continuously with $V$ (as would be implied
from the 2-slit scenario and Eq. (\ref{T2slit})). Instead, $\beta$
remained fixed between resonances, with only discrete jumps by
$+\pi$ (near a resonance) or by $-\pi$ (between resonances). These
discrete jumps are definitely different from the behavior of the
intrinsic phase $\alpha_{QD}$, as shown e.g. in Fig. \ref{ist1}.
Therefore, these experiments cannot be used for direct
measurements of $\alpha_{QD}$, using Eq. (\ref{T2slit}) or Eq.
(\ref{fit}).

The reason for this discrepancy was soon understood. Both the
experiments by Webb {\it et al.} and by Yacoby {\it et al.} were
done on `{\it closed}' interferometers, which differ significantly
from the two-slit geometry. Unlike the latter, the former require
many reflections of the electron waves from the `forks' connecting
the ring with the leads; there is no way to write a $3 \times 3$
unitary matrix, which contains no reflections in two of the three
channels. Each such reflection adds a term to the interference sum
of amplitudes, and modifies the simple two-slit formula. In fact,
it was already shown by Onsager \cite{onsager,but} that unitarity
(conservation of current) and time reversal symmetry imply that
${\cal G}(\phi)={\cal G}(-\phi)$, and therefore $\beta$ (as well
as $\gamma$ etc.) {\it must} be equal to zero or $\pi$, as
observed by Yacoby {\it et al.}. Given the Onsager relation, it is
clear that the data from the closed interferometer should not be
analyzed using the two-slit formula (\ref{T2slit}). However, we
show below that there exists a more complicated formula, which
contains the many reflections from the `forks', and that this
formula can be used to extract the phase $\alpha_{QD}$ from the
closed interferometer data \cite{prl2}.

\begin{figure}[htb]
\begin{center}
\includegraphics[scale=0.9]{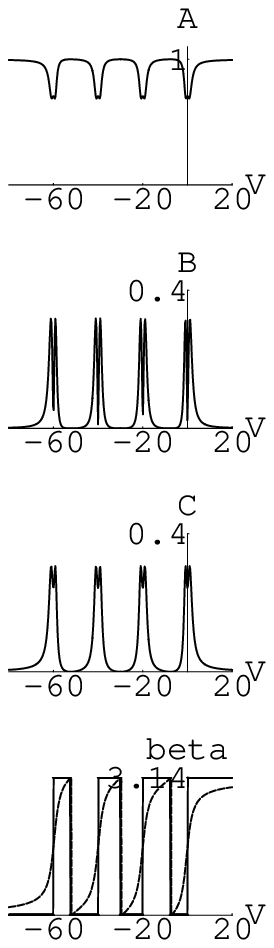}\ \ \ \
\includegraphics[scale=0.9]{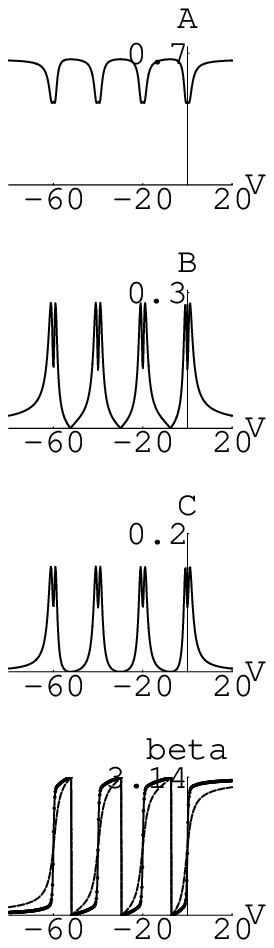}\ \ \ \
\includegraphics[scale=0.9]{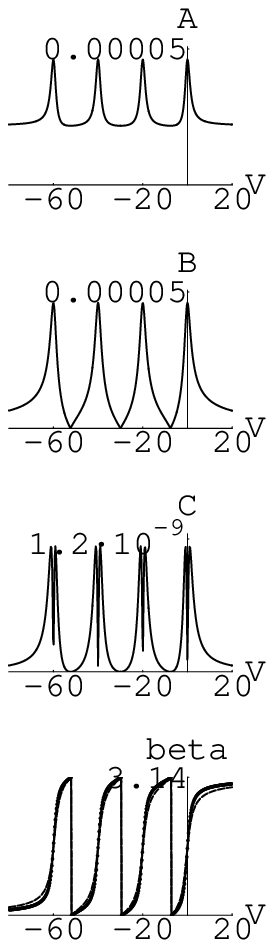}\ \ \ \
\includegraphics[scale=0.9]{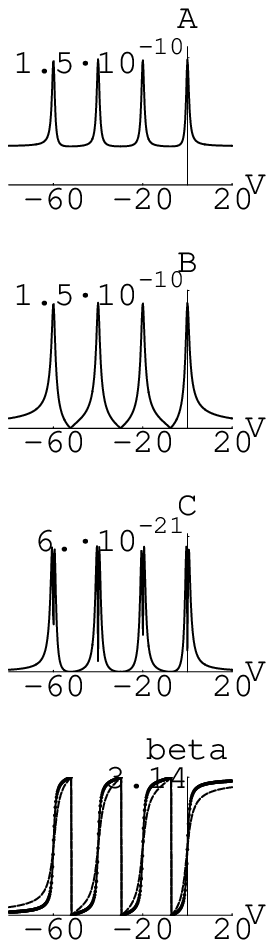}
\end{center}
 \caption{$A,~B,~C$ and $\beta$ for transmission
through the closed AB ring (upper left), and for the open
interferometer with $J_x=.15J$ (upper right) and $J_x=.9J,~1.5J$
(lower left, right). The dashed line shows the exact intrinsic
phase $\alpha_{QD}$, from Fig. \ref{ist1}. For details see Sec.
\ref{sec4} below.} \label{ziur}
\end{figure}

 Later experiments \cite{schuster} opened the interferometer,
using the six-terminal configuration shown schematically in Fig.
\ref{ist2}(b); the additional leads allow losses of electronic
current, thus breaking unitarity. Indeed, fitting the conductance
to Eq. (\ref{T2slit}) yielded a phase $\beta$ which was
qualitatively similar to the calculated $\alpha_{QD}$, as shown in
Fig. \ref{ist1}: a gradual increase through each resonance
(accompanied by peaks in the amplitudes $A$ and $B$), and a sharp
``phase lapse" back to zero between resonances (accompanied by
zeroes in $B$).   These experimental results led to much
theoretical discussion. Some of this \cite{wu,kang} emphasized the
non-trivial effects of the ring itself on the measured results,
even for the closed case. Other theoretical papers
\cite{hack,oreg,ryu,xu,lee,silvestrov,levy} {\it assumed} that the
measured $\beta$ represents the correct $\alpha_{QD}$, and
discussed the possible origins of the observed features, e.g. the
``phase lapse" and the similarity between the data at many
resonances. However, until recently there existed no quantitative
comparison of the measured $\beta$ with the `intrinsic'
$\alpha_{QD}$. In fact, as explained below, it turns out that
$\beta$ {\it depends on the strength of the coupling to the open
channels}: when this coupling vanishes (in the `closed' limit),
$\beta$ jumps between zero and $\pi$. As the coupling increases,
the increase of $\beta$ near a resonance becomes less steep, with
a slope that decreases with increasing coupling \cite{prl1}. Thus,
it is not enough to open the interferometer; one also needs to
choose specific methods of opening, and to tune the relevant
parameters! Below we present a theoretical model, aimed to imitate
the experimental setups of Fig. \ref{ist2}(a) and (b) \cite{bih}.
It has been found that the two-slit conditions can be imitated if
one replaces each lossy channel in Fig. \ref{ist2}(b) by many such
channels, as illustrated in Fig. \ref{ist2}(c). Figure \ref{ziur}
shows examples of our model calculations for $A,~B,~C$ and $\beta$
versus $V$. Qualitatively, these plots look similar to those found
experimentally \cite{yacoby,schuster}. However, as discussed
below, the quantitative results for the open interferometers
depend on details of the opening.

\section{Models for the QD}\label{sec2}

We demonstrate our results for a simple lattice model, in which
the dot is represented by a single site ``$D$" (located at the
origin), on a 1D tight binding chain \cite{ng}. All the on-site
energies are zero, except $\epsilon_D$ on the QD. $\epsilon_D$ can
be varied experimentally by the plunger gate voltage $V$. As usual
for such models, electron-electron interactions are included only
via an on-site Hubbard interaction $U$ on the QD. The hopping
matrix elements $J_{i,i+1}$ on the chain are all equal to $J$,
except on the bonds connected to the QD, where they are
$J_{-1,D}=J_L$ and $J_{D,1}=J_R$. Our Hamiltonian is thus given by
\begin{eqnarray}
{\cal
H}_0=\sum_\sigma\Big(\epsilon_{D}d^{\dagger}_{\sigma}d_{\sigma}+
\frac{U}{2}n_{d\sigma}n_{d\overline{\sigma}} -J\sum_{i \ne
-1,0}[c^{\dagger}_{(i+1)\sigma}c_{i\sigma}+h.c.]-[J_Ld^{\dagger}_{\sigma}c_{-1\sigma}
+J_Rc^{\dagger}_{1\sigma}d_{\sigma}+h.c.]\Bigr ), \label{HH0}
\end{eqnarray}
where $c_{i\sigma}^\dagger$ creates an electron (with spin
$\sigma$) on site $i$. For the unperturbed  chain (with
$\epsilon_D=0$, $U=0$ and $J_L=J_R=J$), one has simple wave
eigenstates, with wave vectors $k$ and eigenenergies
$\epsilon_{k}=-2J\cos ka$ ($a$ is the lattice constant). The
operators on the dot, $d_{\sigma}$ and $d^{ \dagger}_{\sigma}$,
anti-commute with $c_{i\sigma},c^{\dagger}_{i\sigma}$. Also,
$n_{d\sigma}=d^{\dagger}_{\sigma}d_{\sigma}$,
and $\overline{\sigma} \equiv -\sigma$.

Adapting the results of Ref. \cite{ng}, the transmission amplitude
through the QD at zero temperature is given by
\begin{eqnarray}
t_{QD}=-i \gamma_D\sin \alpha_{QD} e^{i\alpha_{QD}} \equiv 2i \sin
|k|a J_LJ_R g_{D}(\epsilon_k)/J, \label{td}
\end{eqnarray}
with the QD asymmetry factor $\gamma_D=2J_LJ_R/(J_L^2+J_R^2)$ and
the ``intrinsic" Green function on the QD,
$g_{D}(\epsilon_k)=1/[\epsilon_k-\epsilon_D-\Sigma_D(\epsilon_k)]$.
Here, $\Sigma_D(\epsilon_k)$ is the self-energy on the QD, which
contains contributions from the leads,
$\Sigma_{D,ext}=-e^{i|k|a}(J_L^2+J_R^2)/J$ (which exists also for
the non-interacting case \cite{prl1}), and from the
electron-electron interactions on the QD itself,
$\Sigma_{D,int}(\omega)$ (which vanishes when $U=0$).
As $\epsilon_D \equiv V$ increases, $\alpha_{QD}$ grows gradually
from zero (far below the resonance), through $\pi/2$ (at the
resonance), towards $\pi$ (far above the resonance).

Interestingly, for this one-dimensional model, normalizing the
measured \begin{eqnarray} {\cal
T}_{QD}=|t_{QD}|^2=\gamma_D^2\sin^2(\alpha_{QD}) \label{sin2}
\end{eqnarray}
by its ($V$-independent) maximum $\max[{\cal T}_{QD}] \equiv
\gamma_D^2$ yields the value of $\alpha_{QD}$. Assuming coherence,
this method for measuring $\alpha_{QD}$ {\it directly} from ${\cal
T}_{QD}$ {\it eliminates the need for any complicated
interferometry}! (However, interferometry is still important,
since it ensures coherence.
 Interestingly, this conclusion holds for {\it
any} Breit-Wigner-like resonance, with an energy-independent
width. It also holds for a multi-level QD, with many resonances).
In the next section we discuss ways of extracting $\alpha_{QD}$
{\it indirectly}, from the {\it closed} AB interferometer
measurements. Comparing results from $\sin^2(\alpha_{QD})={\cal
T}_{QD}/\gamma_D^2 \equiv {\cal T}_{QD}/\max[{\cal T}_{QD}]$, from
the closed interferometer \cite{prl2} and from the open one
\cite{bih} (all with the same QD) should serve as {\em consistency
checks} for this conclusion.

As explained above, at $T=0$ the `intrinsic' transmission
amplitude and phase are directly related to the `bare' Green
function $g_D$ at the Fermi energy, $\epsilon_F$, which is equal
to $\epsilon_k$. Explicit calculations of this Green function, in
the presence of interactions, are non trivial. Although some of
the results below will be given in terms of the full Green
function, it is often useful to use simple expressions to
illustrate specific points. For such purposes, in some of the
explicit calculations below we follow many earlier calculations
\cite{wu,levy,hartzstein,damato,koval}, and ignore the
interactions. For $U=0$, we end up with a simple single-electron
tight-binding model. In this case, the Schr\"odinger wave equation
is written as $(E-\epsilon_i)\psi_i=-\sum_jJ_{ij}\psi_j$, where
the sum is over nearest neighbors of $i$.  The scattering solution
for a wave coming from the left, with wave vector $k$ and energy
$E=-2J\cos ka$, is described by $\psi^L_m=e^{ikam}+re^{-ikam}$ on
the left, and by $\psi^R_m=t e^{ikam}$ on the right. The
calculation of the transmission and reflection amplitudes, $t$ and
$r$, then amounts to solving a finite set of linear equations for
the wave functions inside the scatterer.

Similar linear equations arise for single electron scattering from
more complex geometries, like those shown in Fig. \ref{ist2}. In
each such calculation, we have a scattering element (e.g. the
`ring') connected to two one-dimensional (1D) leads, which have
$J_{i,i+1}=J,~\epsilon_i=0$. All the explicit graphs presented in
the paper are based on the extraction of the total transmission
amplitude $t$ from such equations.

\begin{figure}[htb]
\begin{center}
 \includegraphics[scale=0.6]{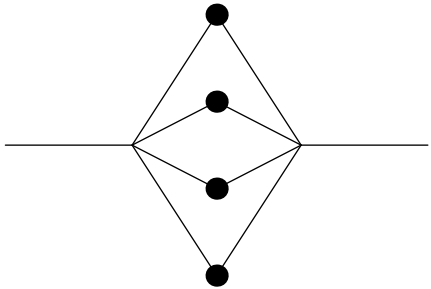}
\end{center}
\caption{Model for a QD with four discrete energy levels.}
\label{qd}
\end{figure}

As discussed above, in many cases one is interested in dots which
have more than one resonance.  Without interactions, it is easy to
generalize the above tight-binding model to a QD with many
discrete energy levels. This is done by a set of smaller dots,
each containing a single resonant state, with energy
$\{\epsilon_{D}=E_R(n),~n=1,...,N \}$. This model is shown in Fig.
\ref{qd} for $N=4$. Each such state (or small dot) is connected to
its left and right nearest neighbors on the leads via bonds with
hopping amplitudes $\{J_L(n),~J_R(n),~n=1,...,N \}$. The QD can
thus be described by $N$ wave functions $\psi_n$, obeying
$[E-E_R(n)]\psi_n=-J_L(n)\psi^L_0-J_R(n)\psi^R_0$ (where we choose
$\psi^L_0=1+r,~\psi^R_0=t$). The exact transmission amplitude is
easily found to be
\begin{equation}
t_{QD}=\frac{S_{LR} 2 i \sin ka}
{(S_{LL}+e^{-ika})(S_{RR}+e^{-ika})-|S_{LR}|^2}, \label{tintr}
\end{equation}
where
\begin{eqnarray}
S_{XY}=\sum_{n}\frac{J_{X}(n)J_Y(n)^\ast}{J[E-E_R(n)]},~~~~X,Y=L,R
\end{eqnarray}
 represent ``bare" Green's functions for sites $L$ and
$R$ (in the previous notation, these were sites $-1$ and $1$ on
the chain).

Figure \ref{ist1} was generated by Eq. (\ref{tintr}), with
equidistant bound state energies, $E_R(n)=V+U(n-1)$. The ``gap"
$U$ can be viewed as the Hartree energy for an electron added to a
QD which already has $n-1$ other electrons \cite{hack}, thus
capturing some aspects of the {\bf Coulomb blockade} behavior of
the scattered electron. We study $t_{QD}$ as function of the
energy $V$, which represents the plunger gate voltage on the QD.
 In this figure and below, we choose $ka=\pi/2$, so
that $E=0$ and the resonances of the transmission, where ${\cal
T}_{QD}=1$, occur exactly when $E_R(n)=E=0$, i.e. when $V=-U(n-1)$
\cite{T0}. Results are not sensitive to $k$ near the band center.
We also use the simple symmetric case, $J_L(n)=J_R(n) \equiv J$,
and measure all energies in units of $J$. As mentioned, this model
reproduces the apparently observed behavior of $\alpha_{QD}$: it
grows smoothly from 0 to $\pi$ as $E$ crosses $E_R(n)$, and
exhibits a sharp ``phase lapse" from $\pi$ to 0 between
neighboring resonances, at points where ${\cal T}_{QD}=0$. These
latter points, associated with zeroes of $S_{LR}$, represent
Fano-like destructive interference between the states on the QD
\cite{fano,ryu,xu,sun,jlt}.

\begin{figure}[htb]
\begin{center}
 \includegraphics[scale=0.6]{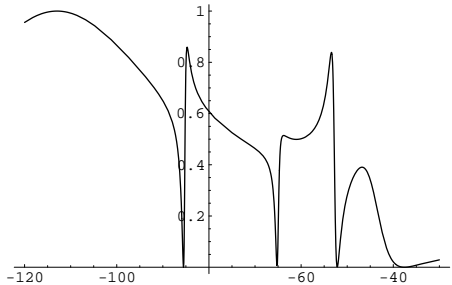}
\end{center}
\caption{Conductance versus gate voltage for a model of a single
electron transistor, based on Eq. (\ref{tintr}).} \label{zgores}
\end{figure}

In fact, Eq. (\ref{tintr}) gives an excellent approximant for the
scattering through a general QD, with several competing
resonances. In Fig. \ref{zgores} we present results for the
transmission through such a QD, with an appropriate
(non-symmetric) choice of the parameters $\{E_R(n),~J_L(n)$ and
$J_R(n)\}$, and $N=5$. This figure reproduces all the experimental
features observed by G\"ores {\it et al.} \cite{gores}, in
scattering from a single electron transistor. Clearly, our Eq.
(\ref{tintr}) gives a much better description of the data, with
less parameters, compared to the sum of individual non-symmetric
Fano expressions \cite{fano} used in Ref. \cite{gores} to fit the
experiments.

 Many earlier theoretical (e.g. \cite{hack}) and experimental
(e.g. \cite{schuster}) papers approximated $t$ by a sum of the
single resonance Breit-Wigner-like (BW) expressions \cite{BW},
\begin{eqnarray}
t_{QD} \approx \sum_n \frac{e^{2ika}2 i \sin ka
J_L(n)J_R(n)^\ast}{E-E_R(n)+e^{ika}[|J_L(n)|^2+|J_R(n)|^2]/J}.
\label{BW}
\end{eqnarray}
Each term here has the form of Eq. (\ref{td}), apart from a
trivial overall phase factor $e^{2ika}$.  Although this form gives
an excellent approximation for $t_{QD}$ near each resonance, it
completely misses the Fano-like zeroes and the ``phase lapses"
between resonances. This happens because the approximation moves
the zeroes off the real energy axis \cite{sun}. As a result, the
approximate $\alpha_{QD}$ never reaches 0 or $\pi$, and exhibits a
smooth decrease from a maximum to a minimum near the correct
``phase lapse" values of $V$.  Since our aim here is to check on
accurate measurements of the ``intrinsic" phase, for a broad range
of the parameters, and since the phase lapse has been a topic of
much recent discussion
\cite{hack,oreg,ryu,xu,lee,silvestrov,levy}, we prefer to use the
exact solutions everywhere. This is particularly important since
typically, available experimental data \cite{schuster} show quite
broad resonances, so that the BW approximation is bound to fail
between them.

We emphasize again: in spite of the close similarity of our
``intrinsic" transmission results with the experiments, the
purpose of this paper is not to relate the calculated $t_{QD}$ to
the experimental systems. This would require a justification for
our choice of the same $J_L(n)$'s and $J_R(n)$'s for all the
resonances, which goes beyond the scope of the present paper.
Rather, we aim to check when the AB interferometer reproduces the
``input" behavior of the QD, by yielding $\beta=\alpha_{QD}$ for
all $V$. If this fails for our simple model then it would surely
fail in the more complicated cases, where electron-electron
interactions (beyond our simple Hartree approximation) become
important \cite{ji}.

\section{Model for the closed AB interferometer}

\begin{figure}[htb]
\begin{center}
 \includegraphics[scale=0.75]{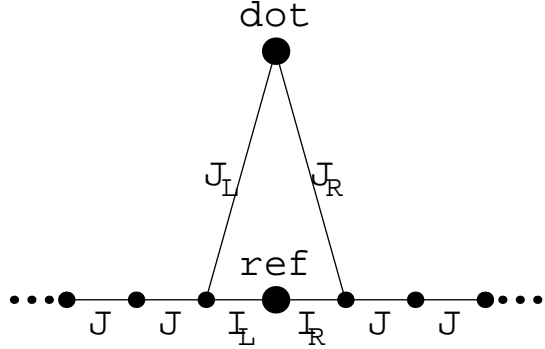}
\end{center}
\caption{Model for the closed AB interferometer.} \label{ABI}
\end{figure}

We next place the above QD on the upper branch of the closed AB
interferometer, as shown in Fig. \ref{ist2}(a).  In the context of
our tight binding model, this translates into the model shown in
Fig. \ref{ABI}: in addition to the path through the QD, we add a
`reference' path, which connects the left and right leads to the
site `ref' via matrix elements $I_L$ and $I_R$. Ignoring electron
interactions on this path, the new Hamiltonian becomes
\begin{eqnarray}
{\cal H}={\cal H}_0+\sum_{\sigma}\Bigl (
\epsilon_{0}c^{\dagger}_{0\sigma}c_{0\sigma}
-I_L[c^{\dagger}_{-1\sigma}c_{0\sigma}+h.c.]
-I_R[c^{\dagger}_{0\sigma}c_{1\sigma}+h.c.]\Bigr ).
\end{eqnarray}
The reference site energy $\epsilon_0$ can be varied
experimentally by an appropriately chosen gate voltage, which we
denote by $V_0$. Adding a magnetic flux $\Phi$ inside the AB ring
now requires adding a phase $\phi$ anywhere around the ring. Using
gauge invariance, we do this by the replacement $J_R \rightarrow
J_Re^{i\phi}$.

 In principle, one can now start from the exact
relation of Eq. (\ref{td}), and add the effects of the `reference'
path perturbatively, as a power series in  $I_L$ and $I_R$. A more
general approach uses the
  standard relation between the $2 \times 2$ scattering matrix
$T_{kk'}$ and the matrix of retarded single-particle Green
functions,
$G_{kk'}(\omega)=\delta_{kk'}g_{k}^0+g^0_kT^\sigma_{kk'}g^0_{k'}$,
with $g^0_{k}(\omega)=1/(\omega-\epsilon_k)$, evaluated on the
energy shell, $\omega=\epsilon_F=\epsilon_k=\epsilon_{k'}$
\cite{hewson}. The equation-of-motion (EOM) method \cite{hewson}
is then used to express $(\omega-\epsilon_k)G_{kk'}(\omega)$ and
$(\omega-\epsilon_k)G_{kd}(\omega)$ as linear combinations of each
other and of $G_D(\omega)$, allowing us to express each of them
(and thus also $t \propto T_{|k|,|k|}$) in terms of the Green
function on the dot, $G_{D}(\omega )$. Since we do not use an
explicit solution for $G_{D}(\omega )$ itself, we don't need to
deal with the higher order correlation functions (due to $U$),
which appear in its EOM.  The result of these procedures has the
form \cite{prl2}
\begin{eqnarray}
t=A_Dt_{QD} e^{i \phi}+A_B t_B, \label{tt1}
\end{eqnarray}
where $A_D=
g_B(\epsilon_k-\epsilon_0)G_D(\epsilon_k)/g_D(\epsilon_k)$ and
 $A_B=1+G_D(\epsilon_k)\Sigma_{ext}(\epsilon_k)$.
 Here, $G_{D}(\omega )=1/[\omega -\epsilon_{D}-\Sigma(\omega )]$
is the fully ``dressed" Green function on the QD, with  the
dressed self-energy $\Sigma=\Sigma_{int}+\Sigma_{ext}$. Both terms
in $\Sigma$ differ from their counterparts in the ``intrinsic"
$\Sigma_D$, by contributions due to the reference path. Also,
\begin{eqnarray}
t_B=-i \gamma_B\sin \delta_B e^{i \delta_B}=2i \sin |k|I_LI_Rg_B/J
\label{tb}
\end{eqnarray}
is the transmission amplitude of the ``background", or
``reference", path (when $J_L=J_R=0$, or $|\epsilon_D| \rightarrow
\infty$), with the bare reference site Green function
$g_B=1/[\epsilon_k-\epsilon_0+e^{i|k|}(I_L^2+I_R^2)/J]$, and the
asymmetry factor $\gamma_B=2I_LI_R/(I_L^2+I_R^2)$.

 Equation (\ref{tt1}) looks like the two-slit formula, Eq.
(\ref{2slit}). However, each of the terms is now {\em
renormalized}: $A_D$ contains all the additional processes in
which the electron ``visits" the reference site ($A_D=1$ when
$I_L=I_R=0$, or when $|\epsilon_0| \rightarrow \infty$), and $A_B$
contains the corrections to $t_B$ due to ``visits" on the dot.  We
now discuss the $\phi$-dependence of ${\cal T} \equiv |t|^2$, in
connection with the Onsager relations and with the possible
indirect extraction of $\alpha_{QD}$.

We first note that
both parts in $\Sigma(\epsilon_k)$ are {\it even} in $\phi$, due
to additive contributions (with equal amplitudes) from clockwise
and counterclockwise motions of the electron around the ring (see
e.g. Refs. \cite{azbel,prl1,hartzstein,W}). In order that ${\cal
T}$ also depends only on $\cos\phi$, as required by the Onsager
relations, the ratio $K\equiv A_Bt_B/(A_Dt_{QD}) \equiv \tilde
x[G_{D}(\epsilon_k)^{-1}+\Sigma_{ext}(\epsilon_k)]$, with the real
coefficient $\tilde x=I_LI_R/[J_LJ_R(\epsilon_k-\epsilon_0)]$,
must be real, i.e.
\begin{eqnarray}
&\Im[G_{D}(\epsilon_k)^{-1}+\Sigma_{ext}(\epsilon_k)] \equiv \Im
\Sigma_{int} \equiv 0. \label{real}
\end{eqnarray}
The same relation follows from the unitarity of the $2 \times 2$
scattering matrix of the ring. This relation already appeared for
the special case of single impurity scattering, in connection with
the Friedel sum rule \cite{langreth}, and was implicitly contained
in Eq. (\ref{td}), where $\Im \Sigma_{D,int}=0$ \cite{ng}.
Equation (\ref{real}) implies that (at $T=0$ and
$\omega=\epsilon_k$) {\it the interaction self-energy}
$\Sigma_{int }(\epsilon_k)$ is {\it real}, and therefore the width
of the resonance, $\Im G_{D}(\epsilon_k)^{-1}$,
is {\it fully determined by the non-interacting self-energy} $\Im
\Sigma_{ext}(\epsilon_k)$.

Since $\Sigma_{ext}(\omega)$ depends only on the (non-interacting)
tight-binding terms, it is easy to calculate it explicitly. We
find
$\Sigma_{ext}(\epsilon_k)=\Sigma_{D,ext}(\epsilon_k)+\Delta_{ext}$,
where
\begin{eqnarray}
\Delta
_{ext}=e^{2i|k|}g_B(J_L^2I_L^2+J_R^2I_R^2+2J_LJ_RI_LI_R\cos\phi)/J^2.
\label{A}
\end{eqnarray}
The  term proportional to $\cos\phi$ comes from the electron
clock- and counterclockwise motion around the AB ``ring".
Similarly, one can write $\Sigma_{int}(\epsilon_k)=\Sigma_{
D,int}(\epsilon_k)+\Delta_{int}$, and thus
$G_{D}(\epsilon_k)^{-1}=g_{D}(\epsilon_k)^{-1}-\Delta$, with
$\Delta=\Delta_{ext}+\Delta_{int}$.  Hence,
$t=A_Dt_D(e^{i\phi}+K)$. Writing also
$A_D=C/[1-g_D(\epsilon_k)\Delta]$, with
$C=(\epsilon_k-\epsilon_0)g_B$, we find
\begin{eqnarray}
{\cal T}=|C|^2{\cal T}_D \frac{1+K^2+2 K\cos\phi}{1-2\Re[g_D
\Delta]+|g_D\Delta|^2}. \label{TT2}
\end{eqnarray}


  Although the numerator in Eq. (\ref{TT2}) looks like the two-slit Eq.
 (\ref{2slit}), with $\beta=0$ or $\pi$ (depending on ${\rm sign} K$),
 the new physics is contained in the denominator -- which
 becomes important in the vicinity of a resonance.
The central term in this denominator depends explicitly on  the
phase of the complex number $g_D$. Since this number is directly
related to $t_{QD}$, via Eq. (\ref{td}), one may expect to extract
$\alpha_{QD}$ from a fit to Eq. (\ref{TT2}), taking advantage of
the dependence of the denominator on $\cos\phi$.
 Physically, this dependence
originates from the infinite sum over electron paths which
circulate the AB ring.
 Ref. \cite{prl2} contains a detailed discussion of the conditions
 for such an extraction.
  Generally, this is not trivial, as one needs
the detailed dependence of $\Delta$ on $\cos\phi$ and on the
various parameters. We have presented this dependence for
$\Delta_{ext}$, but not for $\Delta_{int}$.

The extraction of $\alpha_{QD}$ becomes easy when one may neglect
$\Delta_{int}$. The simplest case for this is for single-electron
scattering, when $\Sigma_{int}=0$. Interactions (i.e. $U \ne 0$)
are also negligible for a relatively {\it open} dot, with small
barriers at its contacts with the leads \cite{mat}.
Another effectively single-electron scattering case arises near a
Coulomb blockade resonance, when the effect of interactions can
simply be absorbed into a Hartree-like shift,
$\epsilon_D+\Sigma_{int} \rightarrow \epsilon_D+N U$, if one {\it
assumes} that $N$ depends smoothly on the number of electrons on
the QD, and not on $\phi$ \cite{W}. If one may neglect
$\Delta_{int}$, then $\Delta \approx \Delta_{ext}$ is given in Eq.
(\ref{A}). Using also Eqs. (\ref{td}) and (\ref{tb}), we find
\begin{eqnarray}
{\cal T}=|C|^2{\cal T}_{QD}\frac{1+K^2+2
K\cos\phi}{1+2P(z+\cos\phi)+Q(z+\cos\phi)^2}, \label{ttt}
\end{eqnarray}
where $z=(J_L^2I_L^2+J_R^2I_R^2)/(2J_LJ_RI_LI_R)$, $P=\Re
[vt_Bt_{QD}]$, $Q=|vt_B|^2{\cal T}_{QD}$, and
$v=e^{2i|k|a}/(2\sin^2|k|a)$ depends only on the Fermi wavevector
$k$, independent of any detail of the interferometer. A
5-parameter fit to the explicit $\phi$-dependence in Eq.
(\ref{ttt}) for given values of $V$ and $V_0$ then yields
$|C|^2{\cal T}_{QD},~K,~z,~P$ and $Q$, and thus
$\cos(\alpha_{QD}+\delta_B+2|k|a)=P/\sqrt{Q}$, from which one can
extract the $V$-dependence of $\alpha_{QD}$. The same
$V$-dependence of $\alpha_{QD}$ is also contained in $K \propto
(\cot\alpha_{QD}+\cot|k|a)$). As discussed after Eq. (\ref{td}),
our model also implies that ${\cal
T}_{QD}=\gamma_D^2\sin^2(\alpha_D)$. Since the $V$-dependence of
${\cal T}_{QD}$ can also be extracted from the fitted values of
either $|C|^2{\cal T}_{QD}$ or $Q$, we end up with several
consistency checks for the determination of $\alpha_{QD}$.
Additional checks arise from direct measurements of ${\cal
T}_{QD}$ and ${\cal T}_B=|t_B|^2$, by taking the limits
$|V_0|=|\epsilon_0| \rightarrow \infty$ or $|V|=|\epsilon_D|
\rightarrow \infty$.

The LHS frame in Fig. \ref{ist6} shows an example of the $V$- and
$\phi$-dependence of ${\cal T}$ for this limit (no interactions),
with $ka=\pi/2$ and $J_L=J_R=I_L=I_R=1,~V_0=4$ (in units of $J$),
implying
 $K=\epsilon_D/\epsilon_0=V/V_0$.  Far away from the resonance
${\cal T} \ll 1$, $Q \ll |P| \ll 1$ and $|K| \gg 1$, yielding the
two-slit-like form  ${\cal T} \approx A+B\cos\phi$, dominated by
its first harmonic, with $B/A \approx 2[K^{-1}-P]$. However, close
the the resonance ${\cal T}$ shows a rich structure; the
denominator in Eq. (\ref{ttt}) generates higher harmonics, and the
two-slit formula is completely wrong. This rich structure may be
missed if one neglects parts of the $\phi$-dependence of $\Delta$,
as done in parts of Ref. \cite{H}. Note also the Fano vanishing
\cite{jlt} of ${\cal T}$ for $V \sim 10$ at $\phi=2n\pi$, with
integer $n$. Without interactions, we can repeat this calculation
for a dot with several resonances, using Eq. ({\ref{tintr}). The
RHS frame in Fig. \ref{ist6} shows results for two resonances,
with $\epsilon_D=\pm 5$. Interestingly, Fig. \ref{ist6} is
qualitatively similar to the experimentally measured transmission
in Ref. \cite{jpn}. However, so far there has been no quantitative
analysis of the experimental data.

\begin{figure}[htb]
\begin{center}
 \includegraphics[scale=1]{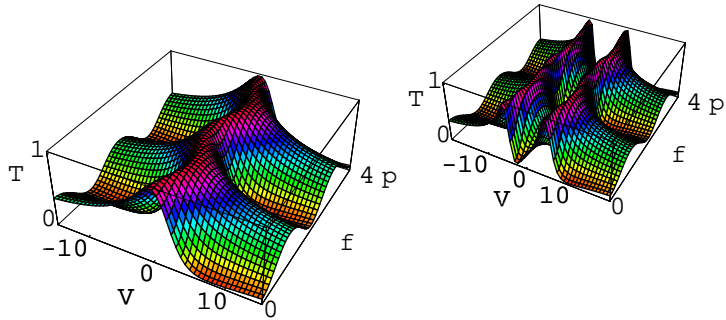}
\end{center}
\caption{AB transmission ${\cal T}$ versus the AB phase $\phi$ and
the gate voltage $V$, for one (LHS) and two (RHS) non-interacting
resonances.} \label{ist6}
\end{figure}

To treat the general case, we need information on $\Delta_{int}$.
First of all, we emphasize that {\it a successful fit to Eq.
(\ref{ttt}) justifies the neglect of the} $\phi$-{\it dependence}
of $\Delta_{int}$. If the various procedures to determine
$\alpha_{QD}$ from Eq. (\ref{ttt}) yield the same $V$-dependence,
this would also confirm that $\Delta_{int}$ is negligibly small. A
failure of this check, or a more complicated dependence of the
measured ${\cal T}$ on $\cos\phi$, would imply that $\Delta_{int}$
is not negligible.

As seen from Eq. (\ref{A}), $\Delta_{ext}$ is fully determined by
a single ``visit" of the electron at ``ref". For small ${\cal
T}_B$, or large $|V_0|=|\epsilon_0|$, it is reasonable to
conjecture that $\Delta_{int}$ is also dominated by such
processes. In that case, we expect $\Delta_{int}$ to be
proportional to the same brackets as in Eq. (\ref{A}), i.e.
$\Delta_{int} \approx w(z+\cos\phi)$, with a real coefficient $w$.
This yields the same dependence of ${\cal T}$ on $\cos\phi$ as in
Eq. (\ref{ttt}), with a shifted coefficient $v$. If $w$ depends
only weakly on $V$, then this shift has little effect on the
determination of $\alpha_{QD}$. Again, the validity of this
approach relies on getting the same $V$-dependence of
$\alpha_{QD}$ from all of its different determinations.

\section{Model for the open AB interferometer}\label{sec4}

Our model for the open interferometer is represented schematically
in Fig. \ref{ist2}(c). In order to obtain explicit expressions,
which are easy to calculate, we again neglect interactions, and
use a simple tight-binding model \cite{bih}. To allow several
leaky branches from each edge of the triangle in Fig. \ref{ABI},
we first generalize the closed interferometer model, and replace
each such edge $s$ by a 1D tight binding model of $M_s$ sites,
with $\epsilon_i=0$ and $J_{i,i+1}=J_{s}$ ($s=\ell,~r,~d$ for the
left and right upper segments and for the lower path,
respectively). Taking advantage of gauge invariance, we attach the
AB phase factor $e^{i\phi}$ to the hopping amplitude from the
right hand ``fork" onto its nearest neighbor on branch $r$, which
we write as $J_r e^{-i\phi}$. Writing the wave functions in
segment $s$ as $\psi^s_m=A_s \eta_s^m+B_s \eta_s^{-m}$, with
$\eta_s$ given by $E=-J_s(\eta_s+\eta_s^{-1})$, it is easy to
express the total transmission and reflection amplitudes through
the interferometer, $t$ and $r$, in terms of the six amplitudes
$\{A_s,~B_s \}$, and obtain six linear equations whose
coefficients also contain $\{ S_{XY} \}$. Having solved these
equations, one finally finds the total transmission amplitude $t$.
Interestingly, the dependence of ${\cal T}$ on $\phi$ for the
closed interferometer remains of the form given in Eq.
(\ref{ttt}). To obtain the LHS frame in Fig. \ref{ziur}, we used
$M_\ell=M_r=6,~M_d=12$, and $J_s=J$. A fit to Eq. (\ref{fit})
indeed gives that $\beta$ jumps between 0 and $\pi$, as in Yacoby
{\it et al.}'s experiments \cite{yacoby}.

We next proceed to model the open interferometer. Pursuing one
possible scenario \cite{prl1}, we model the ``leaking" from each
of the three segments on the ``ring" (imitating the additional
four terminals in the experiment, Fig. \ref{ist2}(b)) by
connecting each site on the three ring segments to a 1D lead,
which allows only an outgoing current to an absorbing reservoir
(Fig. \ref{ist2}(c)). Each such segment is thus replaced by a
``comb" of absorbing ``teeth".

We start by investigating the properties of a single ``comb". The
``base" of the ``comb" is described by a chain of $M$
tight-binding sites, with $J_{m,m+1}=J_c$ and $\epsilon_m=0$. Each
``tooth" is represented by a 1D tight-binding chain, with
$\epsilon_j=0$. The first bond on the ``tooth" has $J_{m,0}=J_x$,
while $J_{j,j+1}=J$ for $j \ge 0$. Assuming only outgoing waves on
the teeth, with wave functions $t_x e^{ikaj}$ and energy $E=-2J
\cos ka$, one can eliminate the ``teeth" from the equations. The
wave functions on the ``base" of the comb are then given by
$\psi^c_m=A_c \eta_c^m+B_c \eta_c^{-m}$, where $\eta_c$ is a
solution of the (complex energy) equation $E+J_x^2
e^{ika}/J=-J_c(\eta_c+\eta_c^{-1})$. When this ``comb" is treated
as our basic scatterer, i.e. connected via $J_{in}$ and $J_{out}$
to our ``standard" two leads, then the transmission and reflection
amplitudes via the ``comb" are given (up to unimportant phases) by
$t=J_{out}(A_c \eta_c^N+B_c/\eta_c^N)/J$ and $r=J_{in}(A_c
\eta_c+B_c/\eta_c)/J-e^{ika}$, and one ends up with two linear
equations for $A_c$ and $B_c$. The results for ${\cal T}=|t|^2$
and ${\cal R}=|r|^2$ are shown, for three values of $M$, in Fig.
\ref{comb}, as functions of $ka \in [0,\pi]$ in the free electron
energy band, for $J_x=.7J$ (left), and as functions of $J_x$, for
$ka=\pi/2$ (right). In the figure, $J_c=J_{in}=J_{out}=J$. It is
rewarding to observe that both ${\cal T}$ and ${\cal R}$ are
almost independent of the electron energy $E$ over a broad range
near the band center. It is also interesting to note that for
these parameters, ${\cal T}$ decreases with $J_x$, but ${\cal R}$
increases with $J_x$. For fixed $J_x$, ${\cal T}$ and ${\cal R}$
exhibit some even-odd oscillations with $M$, but basically ${\cal
T}$ decreases with $M$ while ${\cal R}$ increases towards an
almost constant value for $M>6$. This is understandable: a strong
coupling to the ``teeth" causes a strong decay of the wave
function along the ``comb". Thus, for each value of $M$ one can
find an intermediate optimal region in which both ${\cal T}$ and
${\cal R}$ are small. This region broadens, and has smaller ${\cal
T}$ and ${\cal R}$, for larger $M$.

\begin{figure}[htb]
\begin{center}
 \includegraphics[scale=0.75]{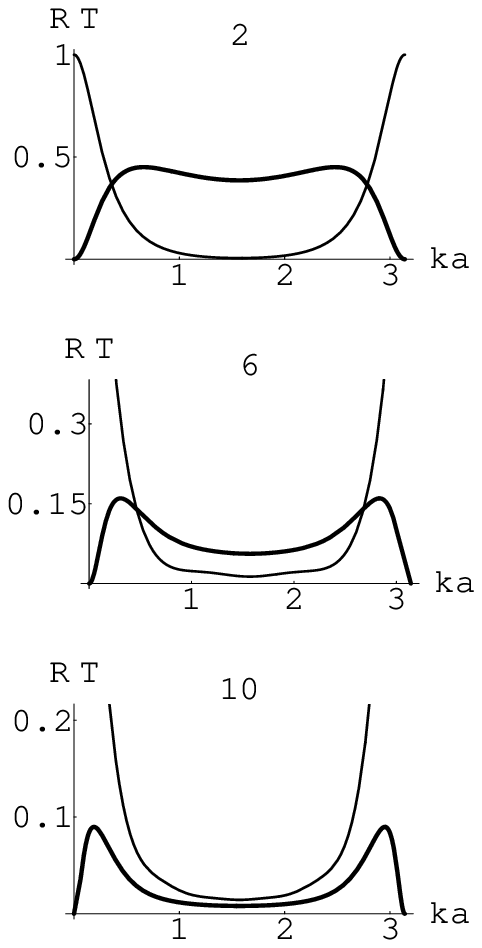}\ \ \ \ \ \includegraphics[scale=0.75]{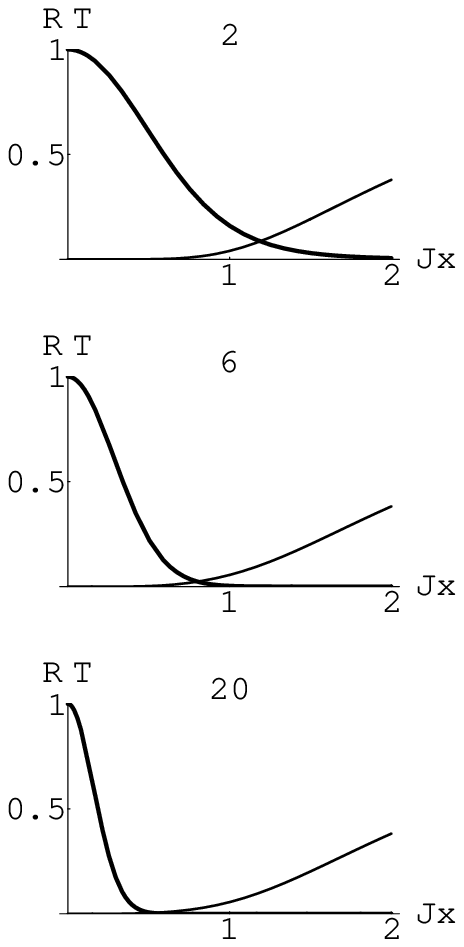}
\end{center}
\caption{Transmission (thick line) and reflection (thin line)
through a ``comb", versus $ka$ at $J_x=.7J$ (left) and versus
$J_x$ at $ka=\pi/2$ (right). The number on each frame gives the
number of ``teeth", $M$.} \label{comb}
\end{figure}

We next place three such ``combs" on the AB interferometer, as in
Fig. \ref{ist2}(c), and study the AB transmission ${\cal T}$ as
function of the various parameters (for the present purposes, the
site ``ref" is just equivalent to the other sites on the lower
edge, i.e. $\epsilon_0=0$). For simplicity, we set the same
parameters for all the combs, and vary the coupling strength
$J_x$. Since each ``tooth" of the ``comb" can be replaced by
adding the complex number $J_x^2e^{ika}/J$ to the energy $E$ in
the equations for $\psi^s_m$ on the ring segments, the mathematics
is similar of that of the ``bare" closed interferometer. The main
difference in the results is that now $\eta_c$ is complex,
yielding a decay of the wave function through each comb. This also
turns the ratio $K$ complex, so that the numerator in Eq.
(\ref{ttt}) must be replaced by $|1+Ke^{i\phi}|^2$, yielding
non-trivial values for $\beta$. To demonstrate qualitative
results, we again choose $M_\ell=M_r=6,~M_d=12$, use
$J_\ell=J_r=J_d=J_c=J$ and keep $ka=\pi/2$ and the QD parameters
$J_L(n)=J_R(n)=J,~N=4,~U=20J$. The choice for the ``comb"
parameters ensures that $A$ and $B$ in Eq. (\ref{fit}) are of the
same order. Other choices give similar qualitative results. Figure
\ref{ziur} shows results for $A,~B,~C$ and $\beta$ as function of
$V$, for several values of $J_x$. Clearly, $J_x=.15J$ gives a
phase $\beta$ which is intermediate between the Onsager jumps of
the left Fig. \ref{ziur} and the exact intrinsic $\alpha_{QD}$ of
Fig. \ref{ist1}. Increasing $J_x$ yields a saturation of $\beta$
onto $\alpha_{QD}$, which persists for a broad range between
$J_x=.5J$ and $J_x=.9J$. However, larger values of $J_x$, e.g.
$J_x=1.5J$, cause a deviation of $\beta$ from $\alpha_{QD}$, due
to the increase of the reflection from each ``comb".
Interestingly, this deviation is {\bf in the same direction} as
for small $J_x$! The reason for this is clear: as the reflection
of each comb increases, the electron ``rattles" in and out of the
QD. This localizes it on the QD, and reduces the width of the QD
resonances. For these large values of $J_x$, one has $|P|, Q \ll
1$ in Eq. (\ref{ttt}). Thus, the two-slit conditions hold, and one
has $B \propto |t_1|$ and $\beta=\alpha_1$. We have solved the
equations for the transmission through the upper branch only
(disconnecting the lower branch altogether), and found that
indeed, the coefficient $c$ in $t_1=c t_{QD}$ is a constant as
long as the reflection of the combs is small. However, as $J_x$
increases above about $.9J$, $c$ is no longer a constant. The
narrower resonances shown in Fig. \ref{ziur} (right) fully agree
with this modified upper branch transmission. In any case,
``optimal combs", with small ${\cal T}$ {\it and} ${\cal R}$, do
yield $\beta=\alpha_{QD}$.

So far, we assumed {\it no} direct losses from the QD itself. It
is easy to add such losses, by connecting a ``lossy" channel to
each resonant state $n$ \cite{prl1}, similar to the ``teeth" of
our ``combs", with a tunneling amplitude $J_x'$. As before, this
introduces a complex addition $J_x'^2e^{ika}$ to $E-E_R(n)$.
Figure \ref{loss} shows the results for the same parameters as
above, but with $J_x=J_x'=.9J$. Clearly, the new imaginary parts
eliminate the Fano-like zero in $B$, and yield a smooth variation
of $\beta$ near the ``intrinsic phase lapses". Although similar to
the behavior arising in the BW approximation, the present effects
are {\it real}, due to physical breaking of the unitarity on the
QD. It is interesting to note that the data of Ref.
\cite{schuster} show similar (and otherwise unexplained) smooth
features. It is however possible that the latter come from finite
temperature averaging \cite{sun}.

\begin{figure}[htb]
\begin{center}
 \includegraphics[scale=0.9]{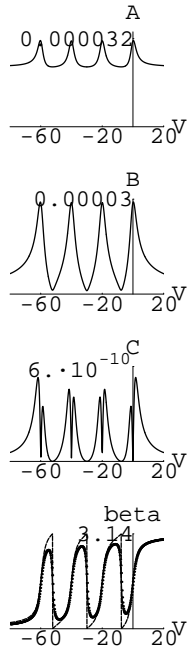}
\end{center}
\caption{Same as Fig. \ref{ziur}, but with a ``lossy" channel
attached to the QD; $J_x=J_x'=.9J$.} \label{loss}
\end{figure}

\section{Concluding remarks}

Basically, we presented three methods to measure the intrinsic
scattering phase of a quantum dot. The first method is based on
Eq. (\ref{sin2}), and does not involve interferometry. The second
is based on Eq. (\ref{TT2}), which allows one to extract
information from measurements on the closed ABI. The third method
uses the open ABI, but requires conditions under which this ABI
behaves as a two-slit interferometer. As stated, a convincing
approach would be to use more than one method, with the same QD,
and to obtain consistent results.

The actual plots shown in this paper were obtained with simple
tight-binding models, without interactions (except for simple
Hartree-like terms in the single electron energy). Therefore,
these plots cannot be used for the strongly interacting case,
particularly in the Kondo regime. Although some aspects of the
interacting case have been included in our analysis for the single
QD or for the closed ABI, the full inclusion of interactions in
practical calculations remain an open problem.

In addition to electron-electron interactions, one might also
consider the effects of other interactions. We have recently
studied the interactions of the electrons with a phonon bath,
which acts only on the QD (still embedded on one path of the
closed ABI). The persistent current $I_p$ around the ``ring", at
steady state, is found to be enhanced in an appropriate range of
the intensity of the acoustic source \cite{persistent}.

\section*{Acknowledgements}
 We thank B. I. Halperin,
M. Heiblum, Y. Levinson, A. Schiller, H. A. Weidenm\"uller and A.
Yacoby for helpful conversations. This project was carried out in
a center of excellence supported by the Israel Science Foundation,
with additional support from the Albert Einstein Minerva Center
for Theoretical Physics at the Weizmann Institute of Science, and
from the German Federal Ministry of Education and Research (BMBF)
within the Framework of the German-Israeli Project Cooperation
(DIP).


\begin{reference}
\bibitem{T0} We assume zero temperature, so that $E$ is equal to
the Fermi energy. Finite temperature replaces the sharp ``phase
lapses" by smooth but fast changes \cite{sun}.
\end{reference}
\end{document}